\documentclass[conference]{IEEEtran}

\IEEEoverridecommandlockouts

\usepackage[pdftex]{graphicx}
\usepackage[cmex10]{amsmath,empheq}
\usepackage{cite}
\usepackage{url}
\usepackage{color}
\usepackage{microtype}
\usepackage{amssymb}
\usepackage{relsize}
\usepackage{lipsum}
\usepackage{epstopdf}
\usepackage{enumitem}
\usepackage{bm}
\usepackage{nicefrac}
\usepackage[percent]{overpic}
\usepackage{subfig}
\usepackage{multirow}
\usepackage{pgfplots}
\usepackage{tikz}
\usepackage{datatool}
\usepackage[autolanguage]{numprint}

\DTLsetseparator{,}

\nprounddigits{1}

\pgfplotsset{compat=1.16}

\begin{document}

\title{Waveform Learning for Reduced Out-of-Band Emissions Under a Nonlinear Power Amplifier}

\newcommand{\todo}[1]{}
\renewcommand{\todo}[1]{{\color{red}{#1}}\PackageWarning{TODO:}{#1!}}
\newcommand{\changed}[1]{{\color{blue}{#1}}}

\author{
\IEEEauthorblockN{Dani Korpi, Mikko Honkala, and Janne M.J. Huttunen}
\IEEEauthorblockA{\textit{Nokia Bell Labs}\\
\textit{Espoo, Finland}}
\and
\IEEEauthorblockN{Fay\c{c}al Ait Aoudia and Jakob Hoydis}
\IEEEauthorblockA{\textit{NVIDIA}\\
\textit{Paris, France}}
\thanks{Fay\c{c}al Ait Aoudia and Jakob Hoydis contributed to this work as members of Nokia Bell Labs until May 2021 and August 2021, respectively.
}
}

\maketitle

\begin{abstract}

Machine learning (ML) has shown great promise in optimizing various aspects of the physical layer processing in wireless communication systems. In this paper, we use ML to learn jointly the transmit waveform and the frequency-domain receiver. In particular, we consider a scenario where the transmitter power amplifier is operating in a nonlinear manner, and ML is used to optimize the waveform to minimize the out-of-band emissions. The system also learns a constellation shape that facilitates pilotless detection by the simultaneously learned receiver. The simulation results show that such an end-to-end optimized system can communicate data more accurately and with less out-of-band emissions than conventional systems, thereby demonstrating the potential of ML in optimizing the air interface. To the best of our knowledge, there are no prior works considering the power amplifier induced emissions in an end-to-end learned system. These findings pave the way towards an ML-native air interface, which could be one of the building blocks of 6G.

\end{abstract}

\section{Introduction}

Implementing digital radio functionality with neural networks is one of the emerging concepts in the field of wireless communications. It is envisioned that the air interface of future 6G radio systems could be built to natively support machine learning (ML). Such ML-native air interface could utilize neural networks in selected parts of the transmitters and receivers in order to improve the system flexibility, spectral efficiency, and robustness, among other things. In particular, learning-based solutions can result in higher performance, for example, under particular channel conditions, hardware impairments, high user equipment (UE) mobility, and/or with very sparse pilot configurations.

In our earlier work, we have considered ML-based physical layer receiver (RX) processing by developing a convolutional neural network (CNN) architecture, which carries out channel estimation, equalization, and demapping jointly. We showed that such a learned receiver can outperform conventional radio receivers in single-input and multiple-output (SIMO) scenarios \cite{Honkala21a}, under spatial multiplexing in a multiple-input and multiple-output (MIMO) receiver \cite{Korpi21b}, and under severe nonlinear distortion due to the transmitter power amplifier (PA) \cite{Pihlajasalo21a}. We have also shown that by considering the radio link in an end-to-end manner, i.e., optimizing the transmitter jointly with the receiver, it is possible to learn a constellation shape that facilitates communication without any pilots \cite{Aoudia21a}. This results in a greatly reduced overhead and, consequently, higher spectral efficiency.

In this paper, we study further the end-to-end optimization of the whole transmission link. In particular, we train the transmitter and receiver to operate under a nonlinear PA with as little out-of-band emissions as possible, while achieving as accurate detection in the receiver side as possible. The former entails learning a waveform more suitable for nonlinearities that results in a better adjacent channel leakage ratio (ACLR), while the latter involves learning jointly the constellation shape and the ML-based receiver, thereby facilitating pilotless transmissions. To the best of our knowledge, there are no prior works addressing the PA-induced out-of-band emissions in an end-to-end learned communication system.

\subsection{Related work}

As mentioned above, there has recently been growing interest into applying different ML techniques in radio physical layer processing. For instance, fully learned receivers for single-input single-output (SISO)/SIMO scenarios have been studied, for example, in \cite{ye18,zhao2018} in addition to our own earlier work in \cite{Honkala21a,Korpi21b,Pihlajasalo21a}. The more extreme approach of learning the complete end-to-end link and thereby optimizing the transmitter and receiver jointly has been studied in \cite{Aoudia21a,oshea17,aoudia20}.

\begin{figure*}
  \centering
  \includegraphics[trim=0px 430px 290px 0px, clip,width=0.9\textwidth]{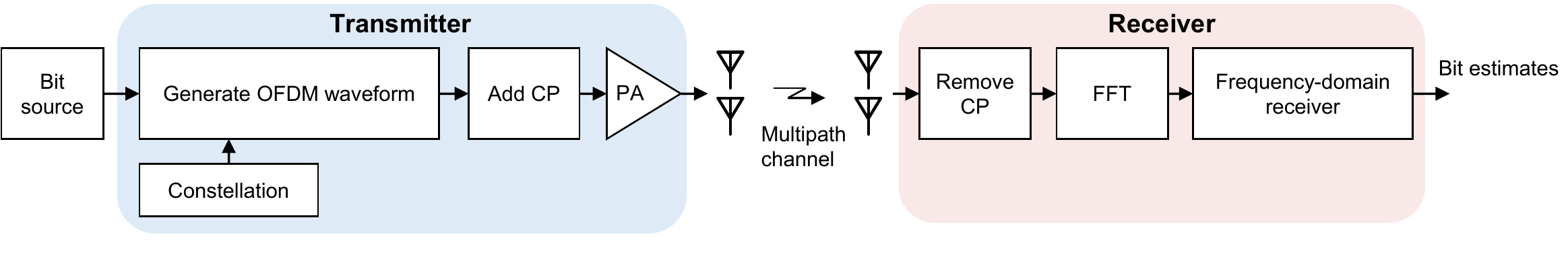}
  \caption{The considered system model, where the transmitter produces modulates the data onto an OFDM waveform and amplifies it with a realistic nonlinear PA. The receiver is assumed to process the received signal in the frequency domain.}
	\label{fig:sys_model}
\end{figure*}

The effects of nonlinearities and ML-based methods for dealing with them have been analyzed, for instance, in \cite{ye18,Felix18,Pihlajasalo21a}. In particular, the work in \cite{ye18} provides some rather preliminary results indicating that an ML-based receiver is capable of dealing with clipping noise, while \cite{Felix18} proposes an end-to-end learned system for communicating efficiently under a nonlinear PA. The work in \cite{Pihlajasalo21a} proposes a hybrid time-frequency domain ML receiver for detecting heavily distorted signals, operating with 5G-compliant waveforms.

\section{System Model}
\label{sec:sys_model}

Figure~\ref{fig:sys_model} depicts the considered baseband-equivalent system model on a high level. In this work, we consider a wireless link based on orthogonal frequency-division multiplexing (OFDM), which is modulated using a given constellation shape. In a conventional transmitter, quadrature amplitude modulation (QAM) is typically used, and this is also assumed in the baseline implementations. After mapping the bits to symbols, they are assigned to the data-carrying resource elements (REs) within the slot. In a conventional transmitter, the demodulation reference signals (DMRS) are also inserted in the frequency domain to their assigned REs. After this, the OFDM symbols within the slot are fed through an inverse Fourier transform (IFFT) to obtain the time-domain waveform. The final procedure before power amplification is to add the cyclic prefix (CP) to the beginning of each OFDM symbol. Then the transmit signal is fed through the PA, after which it propagates through a wireless multipath channel. The noisy and distorted received signal can be written as
\begin{align}
    y(n) = h(n) * \phi \big(x(n)\big) + w(n), \label{eq:rx_signal}
\end{align}
where $h(n)$ denotes the multipath channel response, $*$ is the convolution operation, $\phi \left(\cdot\right)$ is the nonlinear response of the transmitter PA, $x(n)$ is the undistorted transmit waveform, and $w(n)$ is the noise-plus-interference signal. In this work, we assume a memoryless PA model, which means that the nonlinear PA response can be written as
\begin{align}
	\phi \big(x(n)\big) = \sum_{\substack{p = 1 \\p \text{ } \mathrm{odd}}}^{P} f_p \left|x(n)\right|^{p-1} x(n),
\end{align}
where $P$ is the nonlinearity order of the model and $f_p$ denotes the $p$th-order coefficient of the polynomial model.

The first operation in the receiver is the removal of the CP, after which the signal is transformed to the frequency domain with a Fourier transform. In a conventional receiver, this is followed by DMRS-based channel estimation and equalization, after which the receiver calculates the soft bit estimates, often referred to as log-likelihood ratios (LLRs). For further details regarding the conventional frequency-domain receiver processing, please refer to \cite{Korpi21b}.

\section{Proposed End-to-End Learned Communication Link}

\begin{figure*}
  \centering
  \includegraphics[trim=0px 360px 400px 0px, clip,width=0.9\textwidth]{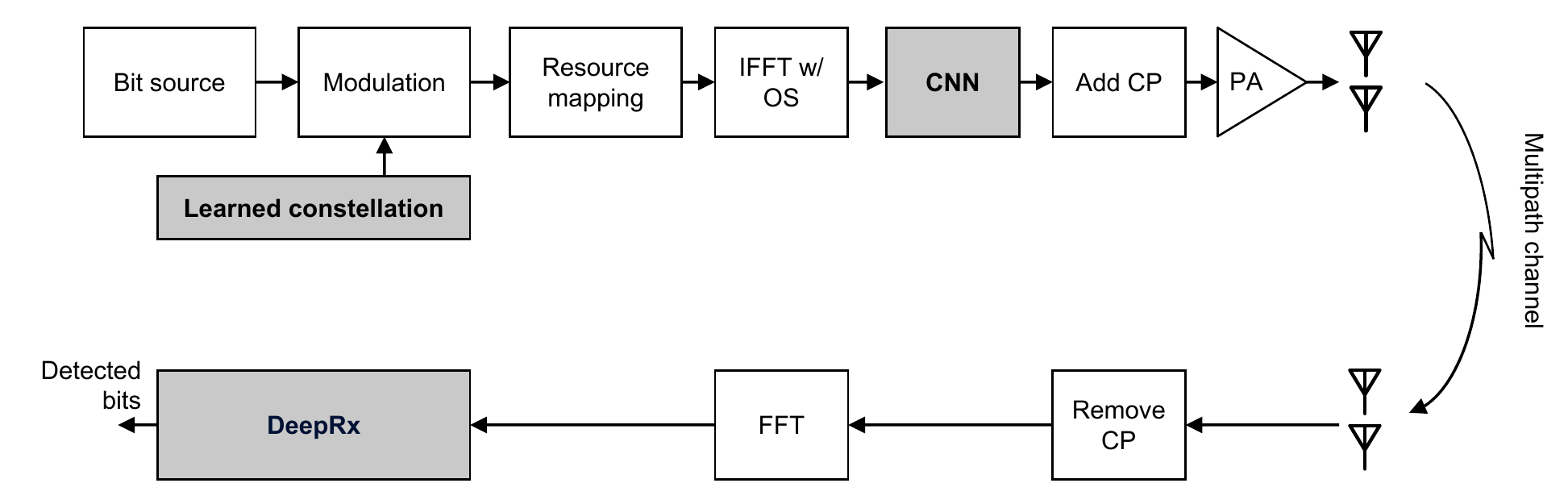}
  \caption{Illustration of the considered end-to-end learned communication link, where the gray blocks are trained jointly.}
	\label{fig:learned_model}
\end{figure*}

Let us then consider the proposed approach of learning the transmit waveform jointly with the receiver. In particular, the objectives of the learning task are the following:
\begin{itemize}
\item Communicate without pilot overhead over a nonlinear PA and a frequency-selective channel;
\item Produce a waveform that results in a lower amount of out-of-band emissions than a conventional OFDM waveform.
\end{itemize}
Figure~\ref{fig:learned_model} depicts the model for the end-to-end learned communication link, which is building on OFDM and assumes a similar slot structure as the conventional system discussed in Section~\ref{sec:sys_model}. The system model illustrates also which parts of the radio link are trained and therefore help in achieving the above objectives. In particular, as opposed to the conventional radio systems, the transmission data is first modulated using a learned constellation, instead of relying on a predefined constellation. That is, the task involves learning two vectors, denoted by $\mathbf{c}_\mathrm{Re} \in \mathbb{R}^{2^{Q_m} \times 1}$ and $\mathbf{c}_\mathrm{Im}  \in \mathbb{R}^{2^{Q_m} \times 1}$, where $2^{Q_m}$ is the size of the constellation and $Q_m$ is the number of bits per symbol. This yields the complex-valued constellation points as $\mathbf{c} = \mathbf{c}_\mathrm{Re} + \jmath \mathbf{c}_\mathrm{Im}$. The actual constellation points are then normalized and centered similar to \cite{Aoudia21a}, which yields the normalized final constellation
\begin{align}
	\bar{\mathbf{c}} = \frac{\mathbf{c} - \frac{1}{2^{Q_m}}\sum_{q=1}^{2^{Q_m}} \mathbf{c}_q}{\sqrt{\frac{1}{2^{Q_m}}\sum_{q=1}^{2^{Q_m}} \left|\mathbf{c}_q \right|^2 - \left|\frac{1}{2^{Q_m}}\sum_{q=1}^{2^{Q_m}} \mathbf{c}_q\right|^2}},
\end{align}
where $\mathbf{c}_q$ refers to the $q$th element of $\mathbf{c}$. Learning the constellation in such a manner allows the link to communicate without additional pilots, as the known (asymmetric) constellation shape can be used by the receiver to detect the distorted symbols and bits

The other learned component in the transmitter side is a CNN operating in the time-domain. The purpose of the CNN is to modify the slightly oversampled transmit waveform in a manner that reduces the out-of-band emissions when fed through the nonlinear PA. In this work we utilize pointwise (1x1) convolutional layers due to the memoryless nature of the PA, and in order to minimize the computational complexity in the TX side. The TX CNN consists of two convolutional layers, and it takes the complex-valued time-domain waveform as its input. The input is first concatenated into a real-valued array, which is fed to the first convolutional layer consisting of 8 filters with hyperbolic tangent as the activation function. The second (output) layer has a linear activation and it outputs the real and imaginary parts of the final transmit waveform, to which the CP is added. As a remark, we note that 1x1 convolutions can be considered as a fully connected network applied separately to each time domain sample. For PAs with memory, the TX CNN should also include convolutional layers with larger filter sizes.

After producing the transmit waveform, effects of nonlinear distortion and wireless multipath channel are then modeled as shown in \eqref{eq:rx_signal}, and the time-domain receiver processing, consisting of CP removal and FFT, is performed in a manner similar to a conventional OFDM receiver. The resulting frequency-domain RX signal is represented by the matrix $\mathbf{Y}\in\mathbb{C}^{N_f \times N_s}$, where $N_f$ is the number of data-carrying subcarriers and $N_s$ is the number of OFDM symbols per slot. This received matrix is then turned into a real-valued array and fed to a DeepRx-type receiver~\cite{Honkala21a}, which is essentially a deep convolutional residual network (ResNet) estimating the received bits jointly over the whole slot. Table~\ref{tab:deeprx_arch} provides a detailed description of the utilized DeepRx architecture. Note that the zero subcarriers stemming from the TX-side oversampling are discarded before DeepRx input. Moreover, since there are no DMRS or pilot symbols in the signal, no raw channel estimate can be calculated and provided as an additional input to the DeepRx, unlike in \cite{Honkala21a}. For a detailed description of an individual ResNet block, please refer to \cite{Honkala21a}.

\begin{table}[!t]
	\renewcommand{\arraystretch}{1.4}
	\caption{The utilized DeepRx CNN ResNet architecture. Note that the ResNet blocks also apply dilations in a similar manner as in \cite{Honkala21a}.} 
	\label{tab:deeprx_arch}
	\centering
	\footnotesize
	\begin{tabular}{|l|p{1.7cm}|l|l|}
	  \hline
		\textbf{Layer} & \textbf{No. of layers / blocks} & \textbf{Filter size} & \textbf{No. of filters}  \\ \hline\hline
		\multicolumn{4}{|l|}{Input $\mathbf{Y}\in\mathbb{C}^{N_f \times N_s}$}\\
		\hline\hline
		\multicolumn{4}{|l|}{Real input $\mathbf{Z}\in\mathbb{R}^{N_f \times N_s \times 2}$}\\ \hline
		Convolutional layer & 1 & (3,3) & 32 \\ \hline
		ResNet Block  & 11 & (3,3) & 32--64 \\ \hline
		Convolutional layer   & 1  & (1,1) & $Q_m$ \\ \hline\hline
		\multicolumn{4}{|l|}{LLR Output $\mathbf{L}^{N_f \times N_s \times Q_m}$}\\ \hline
	\end{tabular}
\end{table}

\subsection*{Training procedure}

As for training the model, the crucial aspect is a properly defined loss function. In this work, the loss function is constructed from two parts: (i) the binary cross entropy (CE) and (ii) the weighted out-of-band emission power in logarithmic scale. The former is obtained by calculating the CE between the transmitted and received bits as
\begin{align}
  \mathrm{CE}_q (\boldsymbol{\theta})\triangleq
 - \frac{1}{B}\sum_{i=1}^{B}\left(b_{iq} \log(\hat{b}_{iq}) + (1 - b_{iq}) \log(1 - \hat{b}_{iq})\right)
\end{align}
where $q$ is the index of the slot, $\hat{b}_{iq}$ is an estimate for the probability that the bit $b_{iq}$ is one, and $B$ is the total number of bits within the slot ($B=N_fN_sQ_m$). Moreover, $\boldsymbol{\theta}$ denotes the trainable parameters of the system (consisting of the parameters of the constellation, TX-side CNN, and DeepRx).

The other loss term is based on the measured out-of-band emissions at PA output. In particular, it is measured from $x_\mathrm{PA}(n) = \phi \big(x(n)\big)$ by calculating the average energy outside the used transmission bandwidth (recall that the TX waveform $x(n)$ is oversampled to facilitate this). In particular, the emission energy is obtained as
\begin{align}
	\mathrm{E}_q (\boldsymbol{\theta}) = \frac{1}{N_s M_\mathrm{OOB}} \sum_{k=1}^{N_s} \sum_{l \in \mathrm{OOB}} \left|X_{klq} \right|^2, \label{eq:E_q}
\end{align}
where $\textrm{OOB}$ contains the indices for the set of unused subcarriers at band edges, $M_\mathrm{OOB}$ is the total number of unused subcarriers, and $X_{klq}$ is the frequency domain representation of $x_\mathrm{PA}(n)$ in the oversampled slot format. It should be noted that in practice this type of emission term includes also artifacts from the oversampling procedure, which in this work is done in the frequency domain. This means that $\mathrm{E}_q (\boldsymbol{\theta})$ contains a small contribution from the IFFT side lobes. However, the impact of this to the total emission power is considerably smaller than that of the nonlinear distortion, meaning that the expression in~\eqref{eq:E_q} is an accurate approximation of the PA-induced out-of-band emissions. Also note that this type of an emission-dependent loss term is essentially equivalent to using an ACLR-dependent loss term, since we always normalize the transmit signal before calculating the emissions. It should also be emphasized that the emissions need to be measured only during training, meaning that no such feedback is required during the inference phase.

With this, the full loss function is obtained by
\begin{align}
  L (\boldsymbol{\theta}) &=\sum_{q=1}^{Q}{\log_2\left(1+{\mathrm{snr}}_q\right)\mathrm{CE}_q (\boldsymbol{\theta})} \nonumber\\
	&+W_E \log \left(\sum_{q=1}^{Q}E_q (\boldsymbol{\theta})\right), \label{eq:loss}
\end{align}
where $Q$ is the batch size, ${\mathrm{snr}}_q$ is the average signal-to-noise ratio of the $q$th slot, and $W_E$ is a weight for the emission loss term. The purpose of the SNR-dependent multiplier is to expedite training of the receiver by giving more weight to the high-SNR data. This prevents the CE term from being dominated by the low-SNR slots, which have inherently less accurate bit estimates. Note that knowledge of the SNR is only needed in the training phase, as the loss is not calculated during inference. Furthermore, it is likely that if the model was trained to operate with SNR-dependent constellation orders, such loss-weighting would be unnecessary.

The training itself is then done by calculating the gradient of the loss $L(\boldsymbol{\theta})$ with respect to the trainable parameters $\boldsymbol{\theta}$ and updating the parameters with stochastic gradient descent (SGD) rule, using a predefined learning rate scheduler. In this work, the so-called Adam optimizer is used \cite{Kingma15a}, which is a popular SGD variant for neural networks.
	
In order to train a hardware-agnostic TX-RX pair, a crucial aspect in training the proposed neural end-to-end link is to randomize also the used PA model. To this end, we use a measured PA response as the basis of the model, and add a random dither term to its polynomial coefficients to get a slightly different nonlinear response for each batch. Moreover, a different set of random PA models is used during training and validation. Altogether, these steps ensure that the trained neural networks have less tendency to specialize to any particular PA response. We would also like to emphasize that in some use-cases it might be desired that the learned link specializes to some particular PA response. Under such a scenario, the training should obviously be carried out using data only from the PA model of interest. We have also experimented with such a training procedure and, as expected, the overall performance is slightly better.

In the results presented below, the PA input power is set to a desired value in order to control the level of nonlinearity, while the PA output signal is then normalized again to unit variance. This ensures that the transmitting neural network does not learn to control out-of-band emissions via backing off the power, but by constructing the transmit waveform in a suitable manner.

\section{Simulation Results}

\begin{figure}
  \centering
  \includegraphics[width=0.3\columnwidth]{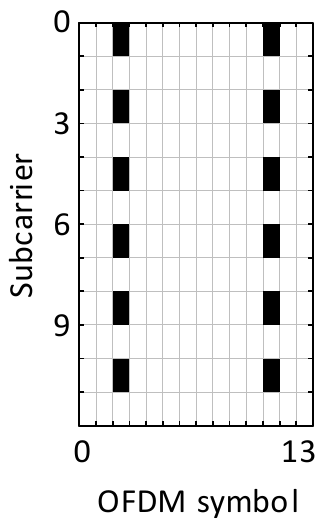}
  \caption{Pilot pattern utilized by the conventional baselines, shown for an individual resource block.}
	\label{fig:pilot_fig}
\end{figure}

\begin{table}
  \setlength{\tabcolsep}{2pt}
    \renewcommand{\arraystretch}{1.3}
    \footnotesize
    \centering
    \caption{Simulation parameters for training and validation.}
    \begin{tabular}{|l|l|l|}
    \hline
    \textbf{Parameter} & \textbf{Value} & \textbf{Randomization}\\
		\hline\hline
		Carrier frequency & 3.5 GHz & None\\
    \hline
		Channel model & UMi LOS, UMi NLOS  & Uniform\\
		\hline
		PA model & Polynomial, $P=17$ & Dithered coefficients \\
		\hline
		UE velocity & 0 m/s -- 25 m/s & Uniform\\
		\hline
		SNR & $0$ dB -- $30$ dB & Uniform\\
		\hline
		Number of subcarriers ($N_f$) & 144 & None\\
		\hline
		TX oversampling factor & 2 & None\\
		\hline
		Subcarrier spacing & 30 kHz & None\\
		\hline
		OFDM symbol duration & 35.7 $\mu$s & None\\
		\hline
		CP duration & 2.7 $\mu$s & None\\
		\hline
		TTI length ($N_s$) & 14 OFDM symbols& None\\
		\hline
		Bits per symbol ($Q_m$) & 6 & None\\
		\hline
    \end{tabular}
    \label{table:param}
  \end{table}
	
In order to evaluate the proposed approach, training and validation data was produced with the simulation parameters shown in Table~\ref{table:param}. The training was carried out for 120~000 iterations with batch size of 100. Each sample within a batch consisted of an individual slot with randomly generated channel response, a random realization of the PA model, random SNR, and random velocity. For validation, a separate randomly generated data was used, where it was ensured that the channel and PA responses were separate from those used during training. The learning rate was first linearly increased from $0$ to the target value within 800 iterations, while it was gradually reduced to zero starting after 36~000 iterations. In our experiments, we utilized target learning rates ranging from $5\cdot 10^{-3}$ to $10^{-2}$ without observing significant differences in the final validation performance.

In the below performance results, two variants of the learned scheme are evaluated: one that includes the TX CNN and is trained with the full loss function defined in \eqref{eq:loss}, and one that does not utilize the TX CNN and is trained using only CE (corresponding to the solution proposed in \cite{Aoudia21a}). This provides insight into the significance of the TX CNN and the loss definition in the proposed scheme. The learned solutions are compared to two conventional benchmark schemes, which are the following:
\begin{itemize}
\item A practical receiver, which estimates the channel with least squares, and interpolates it linearly over the whole slot (extrapolation of the channel estimate beyond the last pilot symbols is done with the nearest neighbor rule). The symbol estimates are obtained with linear minimum mean square error (LMMSE) equalization, after which the soft bits are calculated using the log maximum a-posteriori (log-MAP) rule.
\item A genie-aided receiver, which has perfect channel knowledge, and performs LMMSE equalization followed by log-MAP demapping.
\end{itemize}
Both of these benchmarks utilize a conventional QAM-OFDM waveform and the pilot configuration is as shown in Fig.~\ref{fig:pilot_fig}, where individual pilots are randomly generated quadrature phase shift keying (QPSK) symbols. Note that while the genie-aided receiver does not require the pilots, they are still included in the signal to ensure that the baselines are consistent. Neither receiver assumes any knowledge of the channel error covariance, and no channel coding is utilized in any of the results. Due to the latter, the chosen performance metric in the forthcoming results is the uncoded bit error rate (BER). In our earlier work we have not seen significant difference in the performance gains with or without channel coding \cite{Honkala21a,Korpi21b}.

\newcommand{\aclrBOTEN}{plot_data/aclrs_bo10_mn_d0.csv}
\newcommand{\berBOTEN}{plot_data/bers_bo10_mn_d0.csv}

\newcommand{\constBOTEN}{plot_data/const_bo10_mn_d0.csv}

\newcommand{\aclrBOTENnotdnn}{plot_data/aclrs_bo10_no_tdnn_mn1_b2.csv}
\newcommand{\berBOTENnotdnn}{plot_data/bers_bo10_no_tdnn_mn1_b2.csv}

\newcommand{\aclrBOFF}{plot_data/aclr_boff_v2.csv}
\newcommand{\berBOFF}{plot_data/ber_boff_SNR24_v2.csv}

\DTLloaddb[]{aclr_db}{\aclrBOTEN}
\DTLloaddb[]{aclr_db_notd}{\aclrBOTENnotdnn}

\DTLgetcolumnindex{\mlcol}{aclr_db}{ML}
\DTLgetcolumnindex{\praccol}{aclr_db}{practical}
\DTLgetcolumnindex{\perfcol}{aclr_db}{perfect}

\DTLgetvalue{\MLaclr}{aclr_db}{1}{\mlcol}
\DTLgetvalue{\PRACaclr}{aclr_db}{1}{\praccol}
\DTLgetvalue{\PERFaclr}{aclr_db}{1}{\perfcol}
\DTLgetvalue{\MLaclrnotdd}{aclr_db_notd}{1}{\mlcol}

\begin{figure}[!t]
\begin{tikzpicture}
\begin{axis}[
width=\columnwidth,
height=0.8\columnwidth,
ymode=log,
xmin=0,
xmax=30,
ymin=1e-3,
ymax=1,
grid=both,
xlabel={SNR (dB)},
ylabel={BER},
legend cell align={left},
legend pos=south west,
legend style={font=\scriptsize},
font = \small
]
\addplot[mark=o, mark size=3pt, solid,red,line width=1pt] table[x=SNR, y=practical, col sep=comma] {\berBOTEN};
\addlegendentry{LS, LMMSE (ACLR: \numprint{\PRACaclr}~dB)}
\addplot[mark=square*, mark size=2pt, mark options={solid, fill=blue}, dashed,blue,line width=1pt] table[x=SNR, y=perfect, col sep=comma] {\berBOTEN};
\addlegendentry{Known channel, LMMSE (ACLR: \numprint{\PERFaclr}~dB)}
\addplot[mark=triangle, mark size=3pt, solid,brown,line width=1pt] table[x=SNR, y=ML, col sep=comma] {\berBOTENnotdnn};
\addlegendentry{Learned, w/o TX CNN (ACLR: \numprint{\MLaclrnotdd}~dB)}
\addplot[mark=diamond, mark size=3pt, solid,black,line width=1pt] table[x=SNR, y=ML, col sep=comma] {\berBOTEN};
\addlegendentry{Learned, w/ TX CNN (ACLR: \numprint{\MLaclr}~dB)}
\end{axis}
\end{tikzpicture}
\caption{BER performance of the different solutions when the PA input backoff is 10~dB with respect to unit variance.}
\label{fig:basic_ber}
\end{figure}
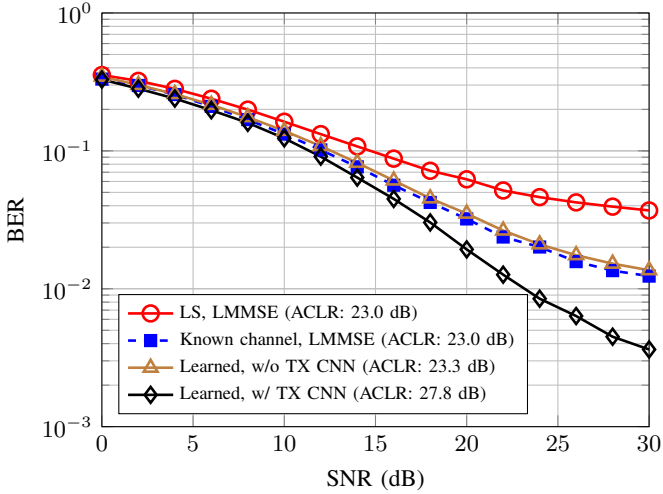

\begin{figure}[!t]
\begin{tikzpicture}
\begin{axis}[
width=\columnwidth,
height=\columnwidth,
xmin=-2,
xmax=2,
ymin=-2,
ymax=2,
grid=both,
font = \small,
xlabel={Real (I)},
ylabel={Imaginary (Q)},
]
\addplot[
		scatter,
    black,
		only marks,
    mark=*,
    mark options={fill=blue, draw opacity=0},
    nodes near coords, 
    point meta=explicit symbolic, 
    every node near coord/.style={anchor=120, font=\tiny} 
    ] table [meta index=2,x=x, y=y, col sep=comma] {\constBOTEN};
\end{axis}
\end{tikzpicture}
\caption{Learned constellation of the proposed scheme with Tx~CNN.}
\label{fig:const}
\end{figure}
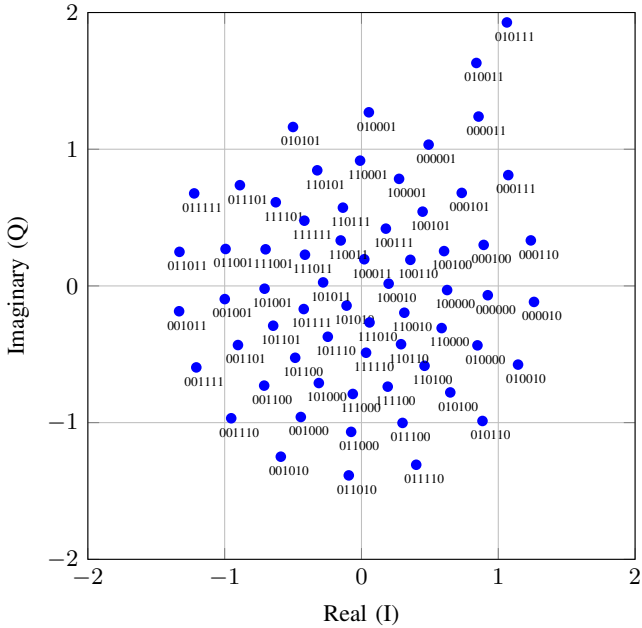

\begin{figure}[!t]
\begin{tikzpicture}
\begin{axis}[
width=\columnwidth,
height=0.8\columnwidth,
ymode=log,
xmin=10,
xmax=25,
ymin=3e-3,
ymax=1e-1,
grid=both,
xlabel={Backoff wrt. unit variance (dB)},
ylabel={BER},
legend cell align={left},
legend pos=north east,
legend style={font=\scriptsize},
font = \small
]
\addplot[mark=o, mark size=3pt, solid,red,line width=1pt] table[x=backoff, y=practical, col sep=comma] {\berBOFF};
\addlegendentry{LS, LMMSE}
\addplot[mark=square*, mark size=2pt, mark options={solid, fill=blue}, dashed,blue,line width=1pt] table[x=backoff, y=perfect, col sep=comma] {\berBOFF};
\addlegendentry{Known channel, LMMSE}
\addplot[mark=triangle, mark size=3pt, solid,brown,line width=1pt] table[x=backoff, y=MLnotdd, col sep=comma] {\berBOFF};
\addlegendentry{Learned, w/o TX CNN}
\addplot[mark=diamond, mark size=3pt, solid,black,line width=1pt] table[x=backoff, y=ML, col sep=comma] {\berBOFF};
\addlegendentry{Learned, w/ TX CNN}
\end{axis}
\end{tikzpicture}
\caption{BER of the different schemes with respect to the PA input backoff when the SNR is fixed at 24~dB}
\label{fig:ber_backoff}
\end{figure}
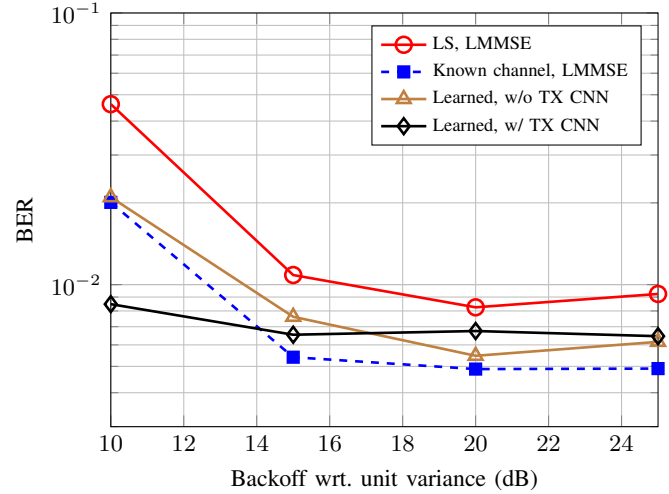

\begin{figure}[!t]
\begin{tikzpicture}
\begin{axis}[
width=\columnwidth,
height=0.8\columnwidth,
xmin=10,
xmax=25,
ymin=20,
ymax=85,
grid=both,
xlabel={Backoff wrt. unit variance (dB)},
ylabel={ACLR (dB)},
legend cell align={left},
legend pos=north west,
legend style={font=\scriptsize},
font = \small
]
\addplot[mark=square*, mark size=2pt, mark options={solid, fill=blue}, dashed,blue,line width=1pt] table[x=backoff, y=perfect, col sep=comma] {\aclrBOFF};
\addlegendentry{OFDM w/ QAM}
\addplot[mark=triangle, mark size=3pt, solid,brown,line width=1pt] table[x=backoff, y=MLnotdd, col sep=comma] {\aclrBOFF};
\addlegendentry{Learned, w/o TX CNN}
\addplot[mark=diamond, mark size=3pt, solid,black,line width=1pt] table[x=backoff, y=ML, col sep=comma] {\aclrBOFF};
\addlegendentry{Learned, w/ TX CNN}
\end{axis}
\end{tikzpicture}
\caption{ACLR of the different schemes with respect to the PA input backoff}
\label{fig:aclr_backoff}
\end{figure}
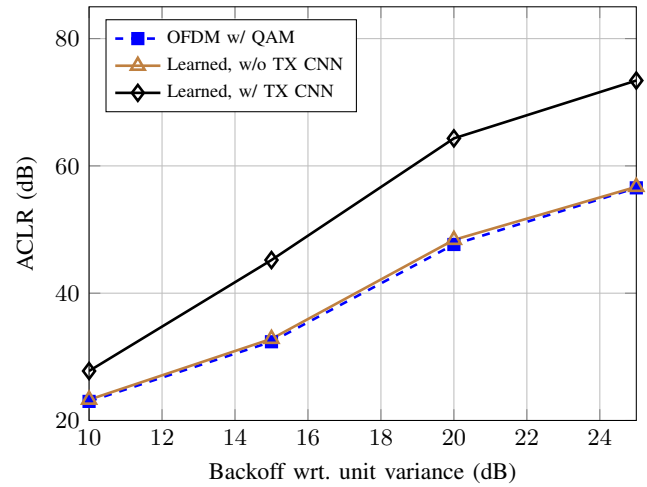

First, Fig.~\ref{fig:basic_ber} shows the BERs under rather severe nonlinear distortion, where the PA input backoff is 10~dB with respect to unit variance (with the utilized PA model, this operating point results in significant nonlinear distortion). The achieved ACLR at the PA output is shown in the legend for each scheme, the ACLR being defined here as the ratio between the inband power and the power on the adjacent bands. The proposed solution with TX CNN and DeepRx receiver achieves a clearly lower BER, while having nearly 5~dB higher ACLR. This highlights the benefits of ML in dealing with hardware impairments on both sides of the radio link. As opposed to this, without the time-domain CNN in the transmitter side, the learned scheme achieves somewhat lower performance, both in terms of emissions and BER. Indeed, now the reduction in out-of-band emissions is essentially negligible. This highlights the benefits of the learned time-domain processing of the transmit waveform and careful formulation of the loss function when dealing with such severe nonlinear distortion.

Another crucial aspect of the proposed solution is the fact that it does not rely on transmitted pilots to detect the signals in the receiver side. Instead, it learns to utilize a learned asymmetric constellation shape to perform purely data-aided detection. Considering this, the proposed solution achieves a clear spectral efficiency gain over the benchmark schemes, both with and without TX-side CNN. To illustrate this, Fig.~\ref{fig:const} shows the learned constellation for the proposed solution (with the TX-side CNN). It can be observed that the constellation contains asymmetric features, which facilitate the DeepRx to correct for the phase and amplitude responses of the unknown channel in a blind manner. Also note that the system learns Gray mapping for the individual constellation points.

In order to analyze the effect of nonlinearity in more detail, Fig.~\ref{fig:ber_backoff} shows the BERs of the different schemes with respect to the PA input backoff when the SNR is fixed at 24~dB. Here higher values of backoff correspond to a more linear PA. With the smallest backoff of 10~dB, the proposed scheme with TX CNN is clearly superior to the other schemes, including the genie-aided LMMSE which has perfect channel knowledge. With higher backoffs, the LMMSE with known channel achieves obviously the lowest BER, but the learned schemes outperform the practical LMMSE baseline throughout the backoff range. However, when the backoff is increased sufficiently, a lower BER is achieved by omitting the TX CNN altogether. This is likely attributed to the fact that the proposed scheme with the TX CNN is still learning to limit the out-of-band emissions, in addition to minimizing CE. This means that its overall learning task is more challenging, which is reflected in the slightly increased BER when the level of nonlinearity is low.

This deduction is confirmed by Fig.~\ref{fig:aclr_backoff} which shows the ACLRs with respect to the PA input backoff, corresponding to the scenario of Fig.~\ref{fig:ber_backoff}. It can be observed that the proposed scheme with TX CNN achieves considerably lower out-of-band emissions compared to the other schemes. In fact, the ACLR gain increases with respect to the backoff, indicating that it is possible to learn to limit the emissions more accurately when the PA is less saturated. This likely depends on the characteristics of the randomized PAs, which are inherently more uniform further away from the area of severe saturation. Altogether, the proposed scheme can fulfill a given ACLR target with roughly 5~dB smaller backoff, which translates to a higher possible TX power. Alternatively, this can also facilitate the use of more power-efficient PA modules.

\section{Conclusion}

In this paper we analyzed an end-to-end learned system operating under a nonlinear power amplifier. In particular, the transmitter was trained to produce a waveform resulting in as little out-of-band emissions as possible, while also learning a constellation shape that facilitates pilotless detection. The transmitter was trained jointly with a DeepRx-type receiver, resulting in a highly optimized link that achieved higher detection accuracy with lower out-of-band emissions than the conventional benchmark schemes. To the best of our knowledge, these are the first findings demonstrating the reduction of power amplifier-induced out-of-band emissions with learned waveforms. This is yet another result showing the benefits of a fully ML-native air interface, which could be a key building block of the future 6G networks.

\section*{Acknowledgments}

This work has been partly funded by the European Commission through the H2020 project Hexa-X (Grant Agreement no. 101015956).

\bibliographystyle{IEEEtran}
\bibliography{IEEEabrv,references}

\begin{thebibliography}{10}
\providecommand{\url}[1]{#1}
\csname url@samestyle\endcsname
\providecommand{\newblock}{\relax}
\providecommand{\bibinfo}[2]{#2}
\providecommand{\BIBentrySTDinterwordspacing}{\spaceskip=0pt\relax}
\providecommand{\BIBentryALTinterwordstretchfactor}{4}
\providecommand{\BIBentryALTinterwordspacing}{\spaceskip=\fontdimen2\font plus
\BIBentryALTinterwordstretchfactor\fontdimen3\font minus
  \fontdimen4\font\relax}
\providecommand{\BIBforeignlanguage}[2]{{%
\expandafter\ifx\csname l@#1\endcsname\relax
\typeout{** WARNING: IEEEtran.bst: No hyphenation pattern has been}%
\typeout{** loaded for the language `#1'. Using the pattern for}%
\typeout{** the default language instead.}%
\else
\language=\csname l@#1\endcsname
\fi
#2}}
\providecommand{\BIBdecl}{\relax}
\BIBdecl

\bibitem{Honkala21a}
M.~Honkala, D.~Korpi, and J.~M.~J. Huttunen, ``{DeepRx}: Fully convolutional
  deep learning receiver,'' \emph{IEEE Transactions on Wireless
  Communications}, vol.~20, no.~6, pp. 3925--3940, Jun. 2021.

\bibitem{Korpi21b}
D.~Korpi, M.~Honkala, J.~M.~J. Huttunen, and V.~Starck, ``{DeepRx} {MIMO}:
  Convolutional {MIMO} detection with learned multiplicative transformations,''
  in \emph{Proc. IEEE International Conference on Communications (ICC)}, Jun.
  2021.

\bibitem{Pihlajasalo21a}
J.~Pihlajasalo, D.~Korpi, M.~Honkala, J.~M.~J. Huttunen, T.~Riihonen,
  J.~Talvitie, A.~Brihuega, M.~A. Uusitalo, and M.~Valkama, ``{HybridDeepRx}:
  Deep learning receiver for high-{EVM} signals,'' in \emph{Proc. IEEE 32nd
  Annual International Symposium on Personal, Indoor and Mobile Radio
  Communications (PIMRC)}, Sep. 2021.

\bibitem{Aoudia21a}
F.~A. Aoudia and J.~Hoydis, ``Trimming the fat from {OFDM}: Pilot- and
  {CP}-less communication with end-to-end learning,'' in \emph{Proc. IEEE
  International Conference on Communications Workshops (ICC Workshops)}, Jun.
  2021.

\bibitem{ye18}
H.~Ye, G.~Y. Li, and B.-H. Juang, ``Power of deep learning for channel
  estimation and signal detection in {OFDM} systems,'' \emph{IEEE
  Communications Letters}, vol.~7, no.~1, pp. 114--117, 2018.

\bibitem{zhao2018}
Z.~Zhao, M.~C. Vuran, F.~Guo, and S.~Scott, ``Deep-waveform: A learned {OFDM}
  receiver based on deep complex convolutional networks,'' 2018.

\bibitem{oshea17}
T.~{O'Shea} and J.~{Hoydis}, ``An introduction to deep learning for the
  physical layer,'' \emph{IEEE Transactions on Cognitive Communications and
  Networking}, vol.~3, no.~4, pp. 563--575, Dec 2017.

\bibitem{aoudia20}
F.~A. Aoudia and J.~Hoydis, ``End-to-end learning for {OFDM}: From neural
  receivers to pilotless communication,'' \emph{arXiv preprint:2009.05261},
  2020.

\bibitem{Felix18}
A.~{Felix}, S.~{Cammerer}, S.~{D\"{o}rner}, J.~{Hoydis}, and S.~{Ten Brink},
  ``{OFDM}-autoencoder for end-to-end learning of communications systems,'' in
  \emph{Proc. IEEE 19th International Workshop on Signal Processing Advances in
  Wireless Communications (SPAWC)}, 2018.

\bibitem{Kingma15a}
D.~Kingma and L.~Ba, ``Adam: A method for stochastic optimization,'' in
  \emph{Proc. International Conference on Learning Representations (ICLR)}, May
  2015.

\end{thebibliography}

\end{document}